\documentclass[preprints,article,inorganics,accept,moreauthors,pdftex]{mdpi}
\usepackage{blkarray}
\firstpage{1} 
\pubvolume{12}
\issuenum{0}
\articlenumber{102}
\pubyear{2024}
\copyrightyear{2024}
\datereceived{29 February 2024} 
\daterevised{26 March 2024} 
\dateaccepted{27 March 2024} 
\datepublished{31 March 2024} 
\hreflink{https://doi.org/10.3390/\linebreak inorganics12040102} 
\pdfoutput=1
\Title{Room-temperature entanglement of the nickel-radical molecular complex (Et$_3$NH)[Ni(hfac)$_2$L] reinforced by the magnetic field}
\TitleCitation{Room-temperature entanglement of the nickel-radical molecular complex (Et$_3$NH)[Ni(hfac)$_2$L]]}

\Author{Jozef Stre\v{c}ka *\orcidA{} and Elham Shahhosseini Shahrabadi \orcidB{}}
\AuthorNames{Jozef Stre\v{c}ka and Elham Shahhosseini Shahrabadi}
\AuthorCitation{Stre\v{c}ka, J.; Shahhosseini, E.}
\address[1]{Department of Theoretical Physics and Astrophysics, Faculty of Science of Pavol Jozef \v{S}af\'{a}rik University in Ko\v{s}ice, 
	Park Angelinum 9, 040~01 Ko\v{s}ice, Slovak republic}
\corres{Correspondence: jozef.strecka@upjs.sk; Tel.: +421-55-234-2276}
\abstract{The bipartite entanglement is comprehensively investigated in the mononuclear molecular complex (Et$_{3}$NH)[Ni(hfac)$_{2}$L], where HL denotes 2-(2-hydroxy-3-methoxy-5-nitrophenyl)-4,4,5,5-tetramethyl-4,5-dihydro-1H-imidazol-3-oxide-1-oxyl and hfacH stands for hexafluoroacetylacetone. From the magnetic point of view, the molecular compound (Et$_{3}$NH)[Ni(hfac)$_{2}$L] consists of an exchange-coupled spin-1 Ni$^{2+}$ magnetic ion and a spin-$\frac{1}{2}$ nitronyl-nitroxide radical substituted nitrophenol. The nickel-radical molecular complex affords an experimental realization of a mixed spin-($\frac{1}{2}$, 1) Heisenberg dimer with a strong antiferromagnetic exchange coupling $J$/$k_{B}$= 505 K and two distinct g-factors g$_{Rad}$=2.005 and g$_{Ni}$=2.275. By adopting this set of magnetic parameters we demonstrate that the Zeeman splitting of a quantum ferrimagnetic ground-state doublet due to a weak magnetic field may substantially reinforce the strength of bipartite entanglement at low temperatures. The molecular compound (Et$_{3}$NH)[Ni(hfac)$_{2}$L] maintains sufficiently strong thermal entanglement even at room temperature, vanishing only above 546 K. Specifically, the thermal entanglement in the nickel-radical molecular complex retains approximately 40\% of the maximum value corresponding to perfectly entangled Bell states at room temperature, which implies that this magnetic compound provides suitable platform of a molecular qubit with potential implications for room-temperature quantum computation and quantum information processing.}
 

\begin{document}
	
\section{Introduction}

The contemporary development of novel quantum technologies requires expertise from various scientific disciplines spanning quantum physics \cite{jaeg18}, materials engineering \cite{bene15} to quantum information science \cite{niel10,pint23}. One of the most challenging tasks in this rapidly evolving and highly interdisciplinary research area is to identify a suitable platform for realizing quantum computers \cite{vinc05}. Although several alternative paradigms are currently being explored to develop novel quantum technologies using diverse physical platforms such as photons, trapped atoms, quantum dots, nuclear and electron spins, or superconducting circuits \cite{ladd10,stre23}, none has yet satisfactorily addressed all fundamental issues related to a practical realization of quantum computation and quantum information processing \cite{vinc05}. The main obstacles in overcoming this intricate problem are associated with the inherent nature of quantum materials, which typically retain their genuine quantum features only at relatively low temperatures or for relatively short coherence times. 

Molecular magnetic materials are compelling candidates for encoding qubits due to their collective spin degrees of freedom, which arise from exchange-coupled electron spins inherent in the individual spin carriers they are composed of \cite{troi11,gait19,atzo19,cruz23}. The main advantage of molecular qubits lies in the capability of targeted design of magnetic molecules with well-controlled number of magnetic centers and interactions between them through a rational bottom-up chemical synthesis \cite{stam09,ferr16}. Another important benefit of  qubits encoded through molecular magnets is the relatively effective decoupling of their electronic degrees of freedom from the environment. This feature allows for tuning their coherence time to be sufficiently long for implementation in quantum computing and quantum information processing \cite{arda07,bade14,zadr15,cruz20}. It has been suggested that  molecular materials with a carefully selected magnetic core may provide a suitable platform for operating as a storage unit of a dynamic random access memory device \cite{leue01}, single- and multiqubit quantum gates \cite{carr07,arda09}, or a quantum battery capable of storing a finite amount of extractable work \cite{cruz22}.

A smart choice of molecular magnetic materials with sufficiently strong exchange coupling constants additionally provides the opportunity to suppress the loss of their genuine quantum features due to thermal effects. The bipartite thermal entanglement may indeed persist up relatively high temperatures as for instance 40~K reported for the spin-1 Ni$^{2+}$ dinuclear complex [Ni$_2$(Medpt)$_2$($\mu$-ox)(H$_2$O)$_2$](ClO$_4$)$_2$ \cite{ghan21}, 50~K for the spin-1 Ni$^{2+}$ tetranuclear complex [Ni$_4$($\mu$-CO$_3$)$_2$(aetpy)$_8$](ClO$_4$)$_4$ \cite{ghan22}, 200~K for the spin-$\frac{1}{2}$ Cu$^{2+}$ coordination polymer Na$_2$Cu$_5$Si$_4$O$_{14}$ \cite{souz08}, 680~K for the spin-$\frac{1}{2}$ Cu$^{2+}$ dinuclear complex [Cu$_2$($\mu$-HCOO)$_4$(HCOO)$_2$]C$_4$H$_{10}$N$_2$ \cite{cruz16} or even 730~K for the spin-$\frac{5}{2}$ Fe$^{3+}$ dinuclear complex [Fe$_2$($\mu_2$-oxo)(C$_3$H$_4$N$_2$)$_6$(C$_2$O$_4$)$_2$] \cite{reis12}. However, none of the aforementioned molecular magnetic materials can be regarded as an experimental realization of the molecular qubit, because they suffer from a limitation of their singlet ground state. 

One possible route towards a physical realization of the molecular qubit is offered by heterobimetallic dinuclear complexes, which represents paradigmatic realizations of the mixed spin-($\frac{1}{2}, S$) Heisenberg dimer with an antiferromagnetic exchange interaction. For this reason, the mixed spin-($\frac{1}{2}, 1$) Heisenberg dimer has recently attracted considerable attention as the simplest member of this class of quantum spin models \cite{guoj07,yang08,ohan15,cenc20,varg21,varg22,nave22,oume23,oume24}. Among other matters, it has been evidenced that the mixed spin-($\frac{1}{2}, 1$) Heisenberg dimer has a higher threshold temperature for thermal entanglement compared to the pure spin-$\frac{1}{2}$ Heisenberg dimer \cite{guoj07}, the strength of thermal entanglement is significantly influenced by a nonuniform magnetic field resulting from different g-factors of individual metal centers \cite{yang08,ohan15}, the Zeeman's splitting of energy levels may substantially enhance the strength of thermal entanglement at low magnetic fields \cite{cenc20,varg21,nave22}, the electric field can serve as another driving force controlling the strength of thermal entanglement \cite{varg22}, and the thermal entanglement can be quite robust against decoherence \cite{oume23,oume24}. From a materials science perspective, the thermal entanglement of experimental realizations of the antiferromagnetic mixed spin-($\frac{1}{2}, 1$) Heisenberg dimer has been studied only for the particular case of the heterodinuclear complex [Ni(dpt)(H$_2$O)Cu(pba)] $\cdot$ 2H$_2$O \cite{hagi99}, which exhibits thermal entanglement up to a threshold temperature of about 140~K comparable to the magnitude of the exchange coupling constant between the spin-$\frac{1}{2}$ Cu$^{2+}$ and spin-1 Ni$^{2+}$ magnetic ions \cite{cenc20,nave22}.  

\begin{figure}[t]
\begin{center}
\includegraphics[width=0.6\textwidth]{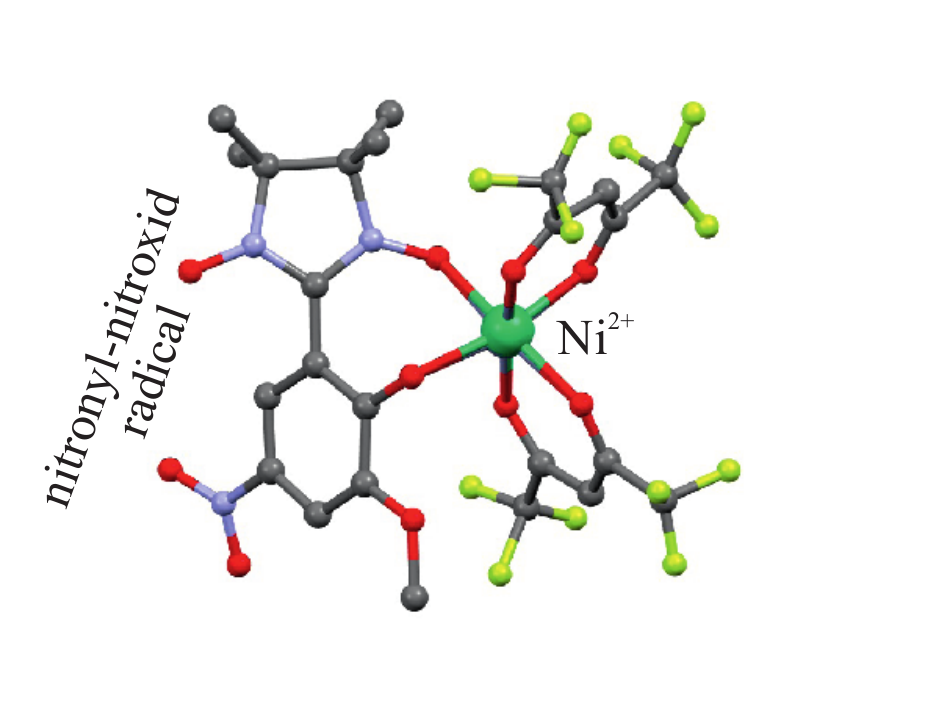}
\end{center}
\vspace*{-0.9cm}
\caption{A magnetic core of the nickel-radical molecular complex (Et$_3$NH)[Ni(hfac)$_2$L] visualized according to crystallographic data deposited at Cambridge Crystallographic Data Centre from reference \cite{spin21}. The used abbreviations: HL = 2-(2-hydroxy-3-methoxy-5-nitrophenyl)-4,4,5,5-tetramethyl-4,5-dihydro-1H-imidazol-3-oxide-1-oxyl and hfacH = hexafluoroacetylacetone.}
\label{fig1}
\end{figure}

In the present paper we will comprehensively examine the strength of bipartite entanglement in the mononuclear molecular complex (Et$_{3}$NH)[Ni(hfac)$_{2}$L] (HL denotes 2-(2-hydroxy-3-methoxy-5-nitrophenyl)-4,4,5,5-tetramethyl-4,5-dihydro-1H-imidazol-3-oxide-1-oxyl and hfacH stands for hexafluoroacetylacetone) \cite{spin21}, which provides completely different platform for an experimental realization of the mixed spin-($\frac{1}{2}, 1$) Heisenberg dimer. Unlike the previous case involving the molecular magnet composed from two distinct metal centers, the magnetic structure of the mononuclear molecular compound (Et$_{3}$NH)[Ni(hfac)$_{2}$L] consists of the spin-$\frac{1}{2}$ nitronyl-nitroxide radical substituted nitrophenol exchange coupled to the spin-1 Ni$^{2+}$ magnetic ion (see figure \ref{fig1}). The main advantage of the molecular qubit encoded in the nickel-radical complex, compared to molecular qubits encoded in mononuclear Cu$^{2+}$ complexes, lies in its greater size, which enables easier control of the qubit state by local magnetic probes such as magnetic force microscopy. It will be demonstrated hereafter that this paradigmatic realization of the mixed spin-($\frac{1}{2}, 1$) Heisenberg dimer may showcase significant thermal entanglement beyond room temperature owing to the sizable exchange interaction $J$/$k_{B}$= 505 K ($k_B$ is Boltzmann's constant) between the spin-$\frac{1}{2}$ nitronyl-nitroxide radical and the spin-1 Ni$^{2+}$ magnetic ion.

The organization of this paper is as follows. In the section \ref{sec:model} we will adapt the mixed spin-($\frac{1}{2}, 1$) Heisenberg dimer for theoretical modeling of the molecular complex 
(Et$_3$NH)[Ni(hfac)$_2$L and describe basic steps of the procedure used for calculating the negativity serving as a measure of the bipartite entanglement. The most interesting results for the quantum and thermal entanglement of the magnetic molecule (Et$_3$NH)[Ni(hfac)$_2$L are then comprehensively discussed in section \ref{sec:result}. Finally, a brief summary of the most important findings and future outlooks are provided in section \ref{sec:conc}.

\section{Model and Method}
\label{sec:model}
In this section, let us first introduce the respective theoretical model of the molecular complex (Et$_3$NH)[Ni(hfac)$_2$L and subsequently describe the basic steps of the procedure allowing calculation of the strength of bipartite thermal entanglement. Taking into consideration an isotropic nature of the exchange coupling between the spin-1 Ni$^{2+}$ magnetic ion and the spin-$\frac{1}{2}$ nitronyl-nitroxide radical, the molecular complex (Et$_3$NH)[Ni(hfac)$_2$L can be theoretically modeled by the mixed spin-($\frac{1}{2}$,1) Heisenberg dimer defined by the Hamiltonian:
\begin{eqnarray}
	\hat{\cal H}= J \hat{\vec{S}} \cdot \hat{\vec{\mu}} -	g_{Rad}\mu_BB\hat{S}^z - g_{Ni}\mu_B B\hat{\mu}^z,
	\label{eq1}
\end{eqnarray} 
where $\hat{\vec{S}} = (\hat{S}^x,\hat{S}^y,\hat{S}^z)$ and $\hat{\vec{\mu}}=(\hat{\mu}^x,\hat{\mu}^y,\hat{\mu}^z)$ denote the spin-$\frac{1}{2}$ and spin-1 operators ascribed to the nitronyl-nitroxide radical and Ni$^{2+}$ magnetic ion, respectively. The coupling constant $J$ determines the isotropic exchange interaction between the spin-$\frac{1}{2}$ nitronyl-nitroxide radical and the spin-1 Ni$^{2+}$ magnetic ion, the parameter $B$ denotes a static external magnetic field, $\mu_B$ is Bohr magneton, while $g_{Rad}$ and $g_{Ni}$ are Land\'e $g$-factors of the nitronyl-nitroxide radical and Ni$^{2+}$ magnetic ion, respectively. In the context of the nickel-radical complex (Et$_{3}$NH)[Ni(hfac)$_{2}$L], the influence of the axial zero-field splitting parameter ($D$) on the spin-1 Ni$^{2+}$ magnetic ion, typically limited at most to a few tens of Kelvins in octahedral environments, can be safely ignored in the Hamiltonian (\ref{eq1}). This is due to the significant exchange coupling $J$/$k_{B}$ = 505 K between the spin-$\frac{1}{2}$ nitronyl-nitroxide radical and the spin-1 Ni$^{2+}$ magnetic ion, which maintains the relative ratio $D/J$ below 2\% and hence, the impact of zero-field splitting on relevant eigenstates becomes negligible \cite{cenc20}.

The Hamiltonian (\ref{eq1}) acquires the following matrix form in the standard basis spanned over the complete set of eigenvectors $\left\vert \varphi_i\right\rangle \!\in\! \left\{ \left\vert\frac{1}{2},1\!\right\rangle,\! \left\vert\frac{1}{2},0\!\right\rangle,\! \left\vert\frac{1}{2},-1 \!\right\rangle,\! \left\vert-\frac{1}{2},1\!\right\rangle,\! \left\vert-\frac{1}{2},0\!\right\rangle,\! \left\vert-\frac{1}{2},-1\!\right\rangle \right\}$ of $z$-components of the spin-$\frac{1}{2}$ and spin-1 operators:
\begin{align}
	&\left\langle \varphi_j \right\vert \hat{\cal H} \left\vert\varphi_i \right\rangle
	=\left(
	\begin{array}{cccccc}
		\frac{J}{2} \!-\! \frac{h_1}{2} \!-\! h_2 & 0 & 0 & 0 & 0 & 0\\
		0 & -\frac{h_1}{2} & 0 & \frac{J}{\sqrt{2}} & 0 & 0\\
		0 & 0 & -\frac{J}{2} \!-\! \frac{h_1}{2} \!+\!h_2 & 0 & \frac{J}{\sqrt{2}} & 0\\
		0 & \frac{J}{\sqrt{2}} & 0 & -\frac{J}{2} \!+\! \frac{h_1}{2} \!-\! h_2 & 0 & 0\\
		0 & 0 & \frac{J}{\sqrt{2}} & 0 & \frac{h_1}{2} & 0  \\
		0 & 0 & 0 & 0 & 0 & \frac{J}{2} \!+\! \frac{h_1}{2} \!+\! h_2
	\end{array}\right).
	\label{eq1a}
\end{align}
For brevity, we have adapted the following notation $h_{1}\!=\!g_{Rad}\mu_BB$ and $h_{2}\!=\!g_{Ni}\mu_BB$ for 'local' Zeeman's terms 
acting on the spin-$\frac{1}{2}$ and spin-1 magnetic entities, which may be different due to difference of the gyromagnetic g-factors 
the nitronyl-nitroxide radical and Ni$^{2+}$ magnetic ion $g_{Rad} \neq g_{Ni}$. An exact analytical diagonalization of the Hamiltonian matrix (\ref{eq1a}) affords the following set of eigenvalues:  
\begin{align}
	E_{1,2}&=\frac{1}{2}\left[J\!\mp\!(h_1\!+\!2h_2)\right],
	\label{eq2}\\
	E_{3,4}&=-\frac{1}{4}\left(J\!+\!2h_2\right)  \mp\frac{1}{4}
	\sqrt{\left[J\!-\!2(h_1\!-\!h_2)\right]^2\!+\!8J^2},
	\label{eq3}\\
	E_{5,6}&=-\frac{1}{4}\left(J\!-\!2h_2\right) \mp\frac{1}{4}
	\sqrt{\left[J\!+\!2(h_1\!-\!h_2)\right]^2\!+\!8J^2},
	\label{eq4}
\end{align} 
and the respective eigenvectors:
\begin{align}
	\displaystyle
	&\left\vert\psi_{1,2}\right\rangle=\left\vert \mp \frac{1}{2}, \mp 1\right\rangle, 
	\label{eq5}\\
	&\left\vert\psi_{3,4}\right\rangle=c_1^{\mp}\left\vert\frac{1}{2},0\right\rangle \mp c_1^{\pm}\left\vert-\frac{1}{2},1\right\rangle,
	\label{eq6}\\
	&\left\vert\psi_{5,6}\right\rangle=c_2^{\pm}\left\vert\frac{1}{2},-1\right\rangle \mp c_2^{\mp}\left\vert-\frac{1}{2},0\right\rangle.
	\label{eq7}
\end{align} 
The eigenvectors \eqref{eq6} and \eqref{eq7} are expressed in terms of the probability amplitudes:
\begin{align}
	c_1^{\pm}=\frac{1}{\sqrt{2}}\sqrt{1\!\pm\!\frac{J\!-\!2(h_1\!-\!h_2)}{\sqrt{\left[J\!-\!2(h_1\!-\!h_2)\right]^2\!+\!8J^2}}
	}\;, 
	\qquad
	c_2^{\pm}=\frac{1}{\sqrt{2}}\sqrt{1\!\pm\!\frac{J\!+\!2(h_1\!-\!h_2)}{\sqrt{\left[J\!+\!2(h_1\!-\!h_2)\right]^2\!+\!8J^2}}
	}\;.
	\label{eq8}
\end{align} 
\\
The strength of quantum and thermal entanglement in pure and mixed states of the mixed spin-($\frac{1}{2}$,1) Heisenberg dimer can be quantified through the quantity \textit{negativity}, which was introduced by Vidal and Werner \cite{vida02} based on the famous Peres-Horodecki separability criterion \cite{pere96,horo96}:
\begin{eqnarray}
	{\cal N}\!=\!\sum_{i=1}^{6} \frac{1}{2} (|\lambda_i|-\lambda_i).
	\label{eq9}
\end{eqnarray} 
The negativity is defined through the eigenvalues $\lambda_i$ of a partially transposed density matrix $\rho^{T_{1/2}}$ derived from the density matrix $\rho$ upon partial transposition $T_{1/2}$ with respect to the spin-$\frac{1}{2}$ subsystem. The negativity is zero for separable states but becomes non-zero for entangled states. It is quite evident from equation (\ref{eq9}) that the bipartite entanglement exists if at least one eigenvalue $\lambda_i$ of the partially transposed density matrix $\rho^{T_{1/2}}$ becomes negative \cite{pere96,horo96}.

A spectral decomposition, which implements the eigenvalues (\ref{eq2})-(\ref{eq4}) and eigenvectors (\ref{eq5})-(\ref{eq7}) of the mixed spin-(1/2,1) Heisenberg dimer, provides a useful tool for calculation of the density operator $\hat{\rho}$:
\begin{align}
	\hat{\rho}&=\frac{1}{Z}\sum_{i=1}^6 \exp(-\beta E_i) \left\vert\psi_i\right\rangle \left\langle \psi_i\right\vert,
	\label{eq10}
\end{align} 
whereby the partition function $Z = \sum_{i=1}^6 \exp(-\beta E_i)$ acquires the following explicit form:
\begin{align}
	Z = 2	{\rm e}^{-\frac{\beta J}{2}}\cosh\left[\frac{\beta}{2} (h_1\!+\!2h_2) \right]\!
	& + 2 {\rm e}^{\frac{\beta J}{4}+\frac{\beta h_2}{2}} 
	\cosh\left(\frac{\beta}{4}\sqrt{\left[J\!-\!2(h_1\!-\!h_2)\right]^2\!+\!8J^2}\right)  
	\nonumber \\
	& + 2 {\rm e}^{\frac{\beta J}{4}-\frac{\beta h_2}{2}}
	\cosh\left(\frac{\beta}{4}\sqrt{\left[J\!+\!2(h_1\!-\!h_2)\right]^2\!+\!8J^2}\right).
	\label{eq11}
\end{align} 
The matrix representation of the density operator \eqref{eq10} gives the following density matrix: 
\begin{eqnarray}
	{\rho}\!&=&\!\left(
	\begin{array}{cccccc}
		\rho_{11} & 0 & 0 & 0 & 0 & 0\\
		0 & \rho_{22} & 0 & \rho_{24} & 0 & 0 \\
		0 & 0 & \rho_{33} & 0 & \rho_{35} & 0\\
		0 & \rho_{42} & 0 & \rho_{44} & 0 & 0\\
		0 & 0 & \rho_{53}  &0 & \rho_{55} & 0\\
		0 & 0 & 0 & 0 & 0 & \rho_{66} 
	\end{array}
	\right),
	\label{eq12}
\end{eqnarray} 
which is defined through the elements $\rho_{ij}$ explicitly quoted in Appendix~\ref{appendixa}. The partially transposed density matrix ${\rho}^{T_{1/2}}$ obtained after a partial transposition $T_{1/2}$ with respect to the spin-$\frac{1}{2}$ subsystem reads:
\begin{eqnarray}
	{\rho}^{T_{1/2}}\!\!\!&=&\!\!\!\left(
	\begin{array}{cccccc}
		\rho_{11} & 0 & 0 & 0 & \rho_{24} & 0\\
		0 & \rho_{22} & 0 & 0 & 0 & \rho_{35}\\
		0 & 0 & \rho_{33} & 0  & 0& 0\\
		0 & 0 & 0 & \rho_{44} & 0 & 0\\
		\rho_{24} & 0 &  0 & 0 & \rho_{55} & 0\\
		0 & \rho_{35} & 0 & 0 & 0 & \rho_{66} 
	\end{array}
	\right).
	\label{eq13}
\end{eqnarray} 
After a straightforward diagonalization of the partially transposed density matrix (\ref{eq13}) one obtains the following spectrum of eigenvalues:
\begin{eqnarray}
	\lambda_1\!\!\!&=&\!\!\! \rho_{33}, \qquad 	\lambda_2 = \rho_{44},\\
	\lambda_{3,4}\!\!\!&=&\!\!\! \frac{\rho_{22}+\rho_{66}}{2}\!\pm\!\frac{1}{2}\sqrt{(\rho_{22}-\rho_{66})^2+4\rho_{35}^2},  \label{eq14a} \\
	\lambda_{5,6}\!\!\!&=&\!\!\! \frac{\rho_{11}+\rho_{55}}{2}\!\pm\!\frac{1}{2}\sqrt{(\rho_{55}-\rho_{11})^2+4\rho_{24}^2}.
	\label{eq14b}
\end{eqnarray} 
The eigenvalues (\ref{eq14a}) and (\ref{eq14b}) with a negative sign before the square root may become negative, which implies according to the formula \eqref{eq9} the possible existence of bipartite entanglement (nonzero negativity). In the next section we will adapt the theoretical results derived for the mixed spin-($\frac{1}{2}$,1) Heisenberg dimer in order to provide a theoretical modeling of bipartite entanglement emerging within the nickel-radical molecular complex (Et$_3$NH)[Ni(hfac)$_2$L].

	\section{Results and discussions}
	\label{sec:result}
	
Before proceeding to a discussion of the most interesting results, let us first briefly review experimental findings reported previously for the nickel-radical molecular complex (Et$_3$NH)[Ni(hfac)$_2$L] in reference \cite{spin21}. Electron paramagnetic resonance spectra revealed two different signals characterized through g-factors $g_{Rad} = 2.005$ and $g_{Ni} = 2.275$, which can be attributed to the spin-$\frac{1}{2}$ nitronyl-nitroxide radical and the spin-1 Ni$^{2+}$ magnetic ion, respectively. It turned out, moreover, that both Land\'e g-factors are almost completely independent of temperature with only negligible changes of the relevant g-factors on the third decimal place. Of course, the g-value $g_{Ni} = 2.275$ can be regarded as the mean value of g-tensor of the spin-1 Ni$^{2+}$ magnetic ion. Besides, theoretical modeling of the temperature dependence of magnetic susceptibility and the low-temperature magnetization curve unveiled a substantial antiferromagnetic exchange coupling $J/k_B = 505$~K between the spin-$\frac{1}{2}$ nitronyl-nitroxide radical and the spin-1 Ni$^{2+}$ magnetic ion. The paramount finding of significant antiferromagnetic exchange coupling thus establishes the mononuclear complex (Et$_3$NH)[Ni(hfac)$_2$L] as a compelling candidate in magnetic materials research capable of showcasing bipartite entanglement persisting well above room temperature. In the following part, we will adjust the parameters of the  exactly solved mixed spin-($\frac{1}{2}$,1) Heisenberg dimer given by the Hamiltonian (\ref{eq1}) to the previously mentioned values $J/k_B = 505$~K, $g_{Rad} = 2.005$, and $g_{Ni} = 2.275$, which allows us to conduct theoretical modeling of the bipartite entanglement in the molecular compound (Et$_3$NH)[Ni(hfac)$_2$L].

\subsection{Quantum entanglement in the nickel-radical molecular complex (Et$_{3}$NH)[Ni(hfac)$_{2}$L]}
\label{ssec:gs}

First, let us examine the strength of bipartite quantum entanglement in the molecular complex (Et$_3$NH)[Ni(hfac)$_2$L] at absolute zero temperature. To achieve this goal, we need to determine the ground state of the mixed spin-($\frac{1}{2}$,1) Heisenberg dimer and construct the relevant density operator. In the absence of an external magnetic field, the ground state of the mixed spin-($\frac{1}{2}$,1) Heisenberg dimer with an isotropic exchange interaction consists of a two-fold degenerate quantum ferrimagnetic state represented by the eigenvectors:
\begin{eqnarray}
	| {\rm QFI}_{\pm} \rangle = \left\{ \begin{array}{ll} 
		\sqrt{\frac{2}{3}} \, \left| - \frac{1}{2} , 1 \right \rangle - \sqrt{\frac{1}{3}} \, \left| \frac{1}{2} , 0 \right \rangle, \\
		\sqrt{\frac{2}{3}} \, \left| \frac{1}{2} , -1 \right \rangle - \sqrt{\frac{1}{3}} \, \left| -\frac{1}{2} , 0 \right \rangle.
	\end{array} \right. 
	\label{gs0}
\end{eqnarray}
The density operator $\hat{\rho} = \frac{1}{2}(\left\vert {\rm QFI}_{+} \right\rangle \left\langle {\rm QFI}_{+}\right\vert + \left\vert {\rm QFI}_{-} \right\rangle \left\langle {\rm QFI}_{-}\right\vert)$ ascribed to the ferrimagnetic doublet \eqref{gs0} implies a moderately strong zero-temperature value of the negativity ${\cal N} = \frac{1}{3}$, which acquires according to equation (\ref{eq9}) two-thirds of its maximal value ${\cal N}_{max} = \frac{1}{2}$ corresponding to perfectly entangled Bell states. It should be pointed out that the sizable energy gap $\Delta E/k_B = 758$~K between a quantum ferrimagnetic ground-state doublet and a spin-quartet excited state implies that the molecular complex (Et$_3$NH)[Ni(hfac)$_2$L] can be regarded, up to extremely high temperatures and magnetic fields, 
as a compelling experimental realization of the molecular qubit. 

An even more intriguing situation emerges when the external magnetic field lifts the degeneracy of the quantum ferrimagnetic doublet \eqref{gs0} through Zeeman's splitting of energy levels. The Zeeman effect consequently leads to a nondegenerate quantum ferrimagnetic ground state given by the eigenvector:
\begin{eqnarray}
	| {\rm QFI}_{+} \rangle = c_{+} \left| -\frac{1}{2}, 1 \right \rangle - c_{-} \left| \frac{1}{2}, 0 \right \rangle,
	\label{gsf}
\end{eqnarray}
which is characterized through the probability amplitudes depending on a mutual interplay of the exchange-coupling constant $J$ and the magnetic-field term $\mu_B B(g_{Ni}\!-\!g_{Rad})$ 
involving the difference of the relevant g-factors:
\begin{eqnarray}
	c_{\pm} = \frac{1}{\sqrt{2}} \sqrt{1\!\pm\!\frac{J\!+\!2 \mu_B B (g_{Ni}\!-\!g_{Rad})}{\sqrt{\left[J\!+\!2 \mu_B B(g_{Ni}\!-\!g_{Rad})\right]^2\!+\!8J^2}}
	}\;.
	\label{pagsf}
\end{eqnarray}
The assessment of bipartite quantum entanglement within the nondegenerate ferrimagnetic ground state $| {\rm QFI}_{+} \rangle$ given by equations \eqref{gsf}-\eqref{pagsf} can be readily accomplished  with the help of the respective density operator $\hat{\rho} = \left\vert {\rm QFI}_{+} \right\rangle \left\langle {\rm QFI}_{+}\right\vert$ being consistent with the following zero-temperature value of the negativity:
\allowdisplaybreaks
\begin{align}
	{\cal N}&=\frac{\sqrt{2}J}{\sqrt{\left[J+2(g_{Ni}-g_{Rad}) \mu_B B \right]^2 + 8J^2}}.
	\label{eq18}
\end{align} 
A low-field asymptotic value of the negativity ${\cal N} = \frac{\sqrt{2}}{3} \approx 0.47$ calculated according to equation \eqref{eq18} for the nondegenerate ferrimagnetic ground state $|{\rm QFI}_{+} \rangle$ in the limit $B \to 0^{+}$ is indicative of an unexpected enhancement of the bipartite quantum entanglement when compared to the zero-field value ${\cal N} = \frac{1}{3}$ corresponding to the two-fold degenerate ferrimagnetic ground states $|{\rm QFI}_{\pm} \rangle$ emergent strictly at zero magnetic field $B=0$. Note that this phenomenon contradicts the conventional expectation of magnetic-field-induced suppression of the bipartite quantum entanglement, because it suggests instead its significant rise (enhancement approximately by 41 \%) at nonzero magnetic fields due to the Zeeman effect. Although a subsequent rise of the magnetic field causes a gradual reduction in the strength of bipartite quantum entanglement, this effect is quite weak for achievable magnetic-field strengths due to the negligible size of the magnetic-field-dependent term $(g_{Ni}-g_{Rad}) \mu_B B$ relative to the sizable exchange interaction $J/k_B = 505$~K. It will be demonstrated hereafter that the sudden increase in  bipartite quantum entanglement due to the magnetic field persists even at low temperatures what underscores its potential significance for experimental verification.

\subsection{Thermal entanglement in nickel-radical molecular complex (Et$_{3}$NH)[Ni(hfac)$_{2}$L]}
\label{te}

\begin{figure}[t]
\begin{center}
\includegraphics[width=0.75\textwidth]{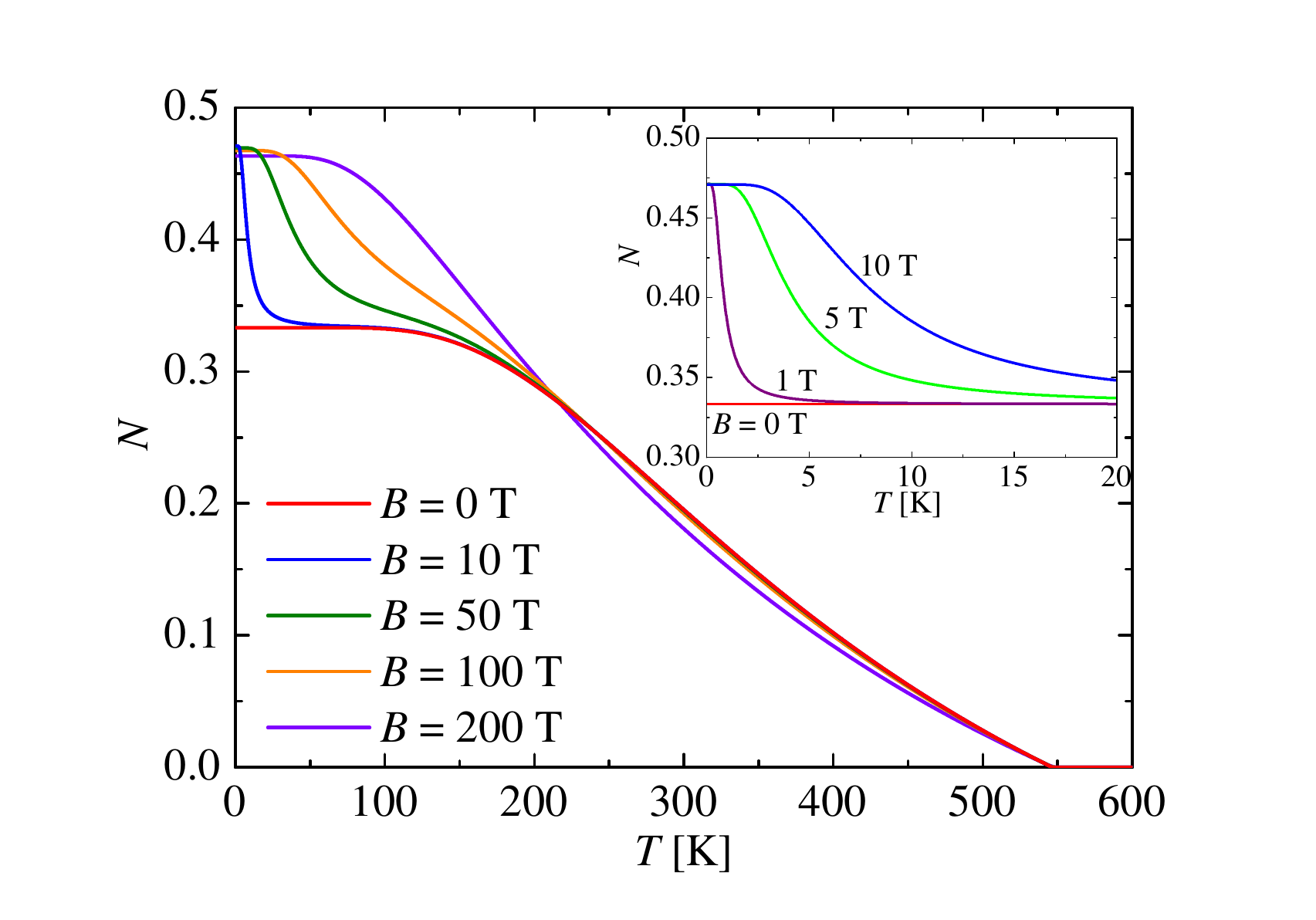}
\end{center}
\vspace*{-0.8cm}
\caption{Temperature variations of the negativity calculated at a few selected values of the external magnetic field for the mixed spin-($\frac{1}{2}$,1) Heisenberg dimer given by the Hamiltonian (\ref{eq1}) with the parameter set $J/k_B = 505$~K, $g_{Rad} = 2.005$, and $g_{Ni} = 2.275$ adjusted to a theoretical modeling of the molecular complex (Et$_3$NH)[Ni(hfac)$_2$L]. The inset shows a detail from the low-temperature region and it included a few additional values of the small magnetic field.}
\label{fig2}
\end{figure}

Next, let us investigate the strength of bipartite thermal entanglement in the molecular complex (Et$_3$NH)[Ni(hfac)$_2$L] at nonzero temperatures. Typical temperature dependencies of the negativity of the mixed spin-($\frac{1}{2}$,1) Heisenberg dimer with the parameter set $J/k_B = 505$~K, $g_{Rad} = 2.005$, and $g_{Ni} = 2.275$ adjusted to a theoretical modeling of the molecular compound (Et$_3$NH)[Ni(hfac)$_2$L] are depicted in Fig. \ref{fig2} for a few selected values of the external magnetic field. In agreement with the previous ground-state analysis, the negativity starts at zero magnetic field from the zero-temperature asymptotic value ${\cal N} = \frac{1}{3}$ and then it shows a gradual temperature-induced decline until the negativity completely vanishes at the threshold temperature $T \approx 546$~K. Contrary to this, it is evident from Fig. \ref{fig2} and its inset that the negativity starts at relatively small magnetic fields ($B \lesssim 50$~T) from the higher zero-temperature asymptotic value ${\cal N} = \frac{\sqrt{2}}{3} \approx 0.47$ in accordance with the previous arguments concerning with the magnetic-field-driven enhancement of bipartite quantum entanglement. The negativity at relatively small magnetic fields subsequently shows a sudden drop upon increasing of temperature until it tends towards the respective zero-field value before the negativity finally displays a gradual temperature-induced decline terminating at the same threshold temperature $T \approx 546$~K as the zero-field case. Note that the initial thermally-driven reduction of the negativity is the steeper, the smaller the external magnetic field is, whereas this phenomenon gradually vanishes at higher magnetic fields. A detectable reduction of zero-temperature value of the negativity ${\cal N} = \frac{\sqrt{2}}{3} \approx 0.47$ due to the external magnetic field occurs merely at extremely high magnetic fields $B \gtrsim 200$~T.

\begin{figure}[h]
\begin{center}
\includegraphics[width=0.75\textwidth]{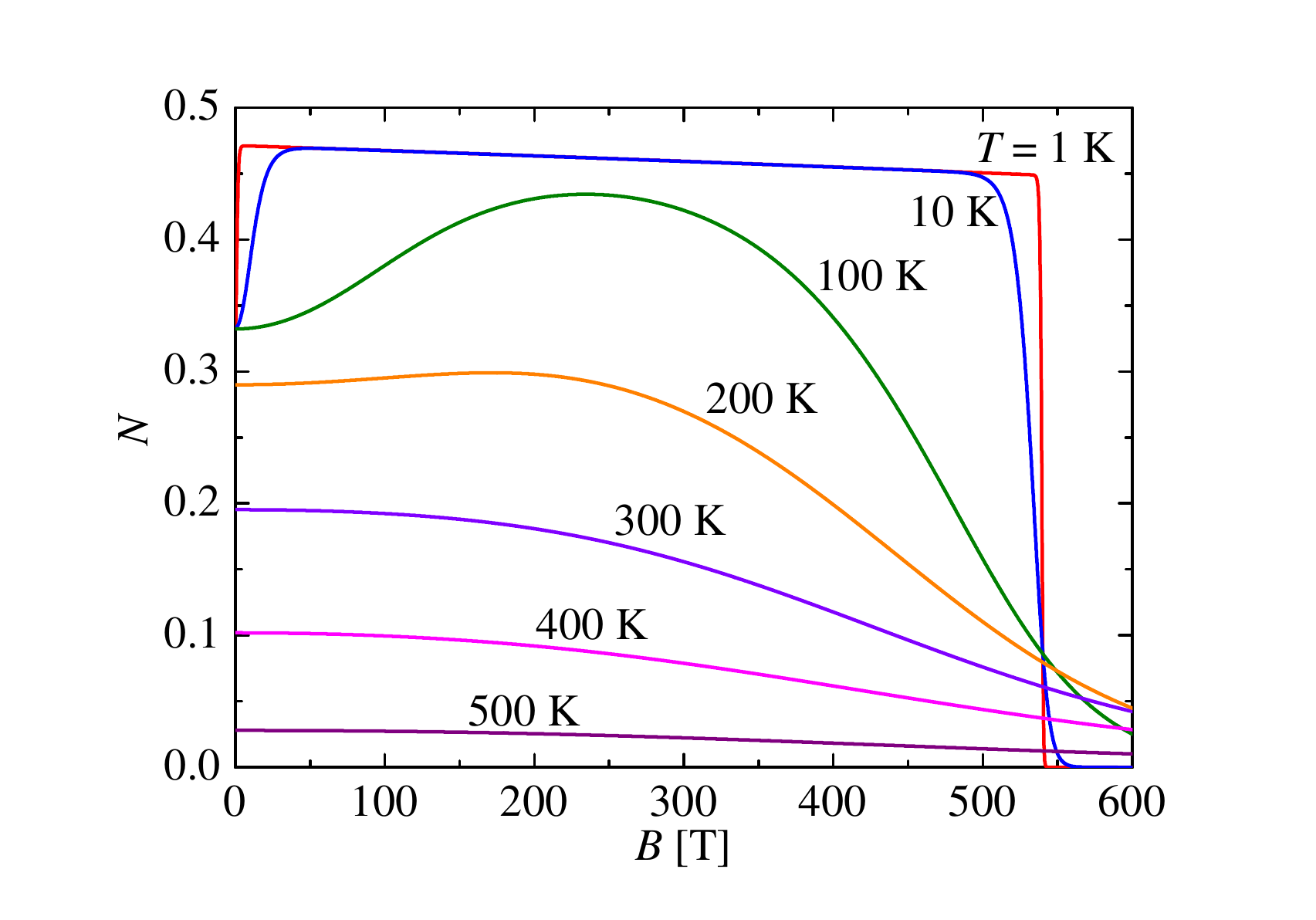}
\end{center}
\vspace*{-0.8cm}
\caption{Magnetic-field-driven changes of the negativity calculated at a few selected values of temperature for the mixed spin-($\frac{1}{2}$,1) Heisenberg dimer given by the Hamiltonian (\ref{eq1}) with the parameter set $J/k_B = 505$~K, $g_{Rad} = 2.005$, and $g_{Ni} = 2.275$ adjusted to a theoretical modeling of the molecular complex (Et$_3$NH)[Ni(hfac)$_2$L].}
\label{fig3}
\end{figure}

Furthermore, a few typical isothermal dependencies of the negativity on a magnetic field are plotted in Fig. \ref{fig3} for the mixed spin-($\frac{1}{2}$,1) Heisenberg dimer with the parameter set $J/k_B = 505$~K, $g_{Rad} = 2.005$, and $g_{Ni} = 2.275$ adjusted to a theoretical modeling of the molecular compound (Et$_3$NH)[Ni(hfac)$_2$L]. At low enough temperatures $T \lesssim 10$~K the negativity shows an abrupt magnetic-field-induced rise from the zero-field value ${\cal N} = \frac{1}{3}$ towards the much higher value ${\cal N} = \frac{\sqrt{2}}{3} \approx 0.47$, which is subsequently followed by a gradual quasi-linear decrease achieved upon further strengthening of the magnetic field until the negativity suddenly drops down to zero at the extremely high magnetic field $B \approx 540$~T. While a small magnetic field significantly strengthens the thermal entanglement roughly by 41\% due to the Zeeman splitting of the ferrimagnetic ground-state doublet, the subsequent reduction of thermal entanglement observable upon further increase of the magnetic field is quite negligible. Namely, the magnetic-field change of 500~T decreases the negativity only by approximately 2\%. It is noteworthy that the unconventional magnetic-field-driven strengthening of thermal entanglement is gradually suppressed upon increasing of temperature as exemplified by the isothermal dependencies of the negativity calculated at temperatures $T = 100$~K and 200~K. At and above room temperature, the negativity displays a conventional monotonous decrease upon increasing the magnetic field, whereby the strength of thermal entanglement is kept almost constant in the magnetic-field range $B \lesssim 10$~T effective for technological applications. From this perspective, it is worthwhile to remark that the strength of bipartite entanglement of the molecular compound (Et$_3$NH)[Ni(hfac)$_2$L] at room temperature acquires roughly 40\% of its maximum value corresponding to the fully entangled Bell states independently of the magnetic-field strengths accessible for contemporary real-world technologies ($B < 10$~T). 

\begin{figure}[t]
\begin{center}
\includegraphics[width=0.75\textwidth]{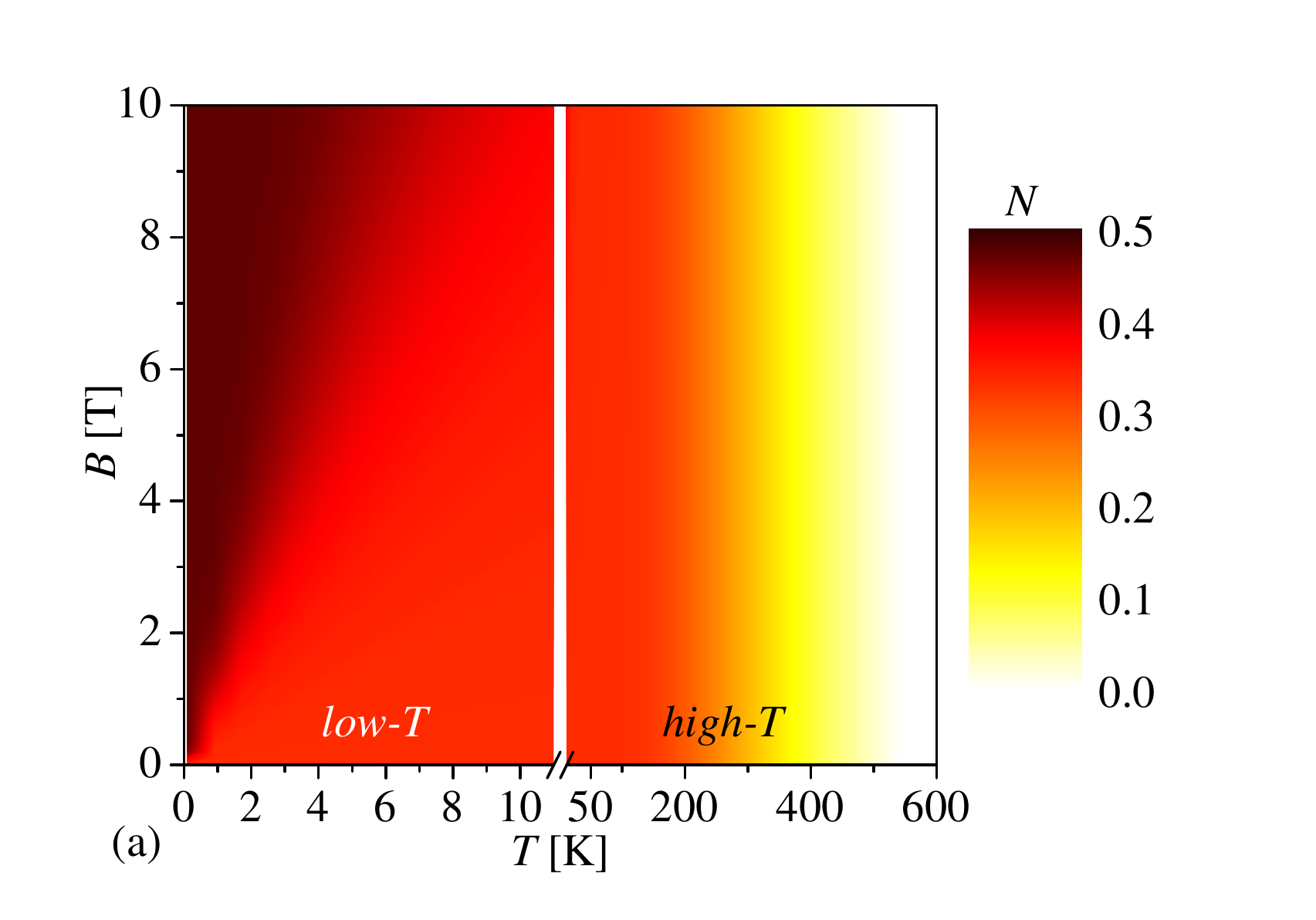}
\includegraphics[width=0.75\textwidth]{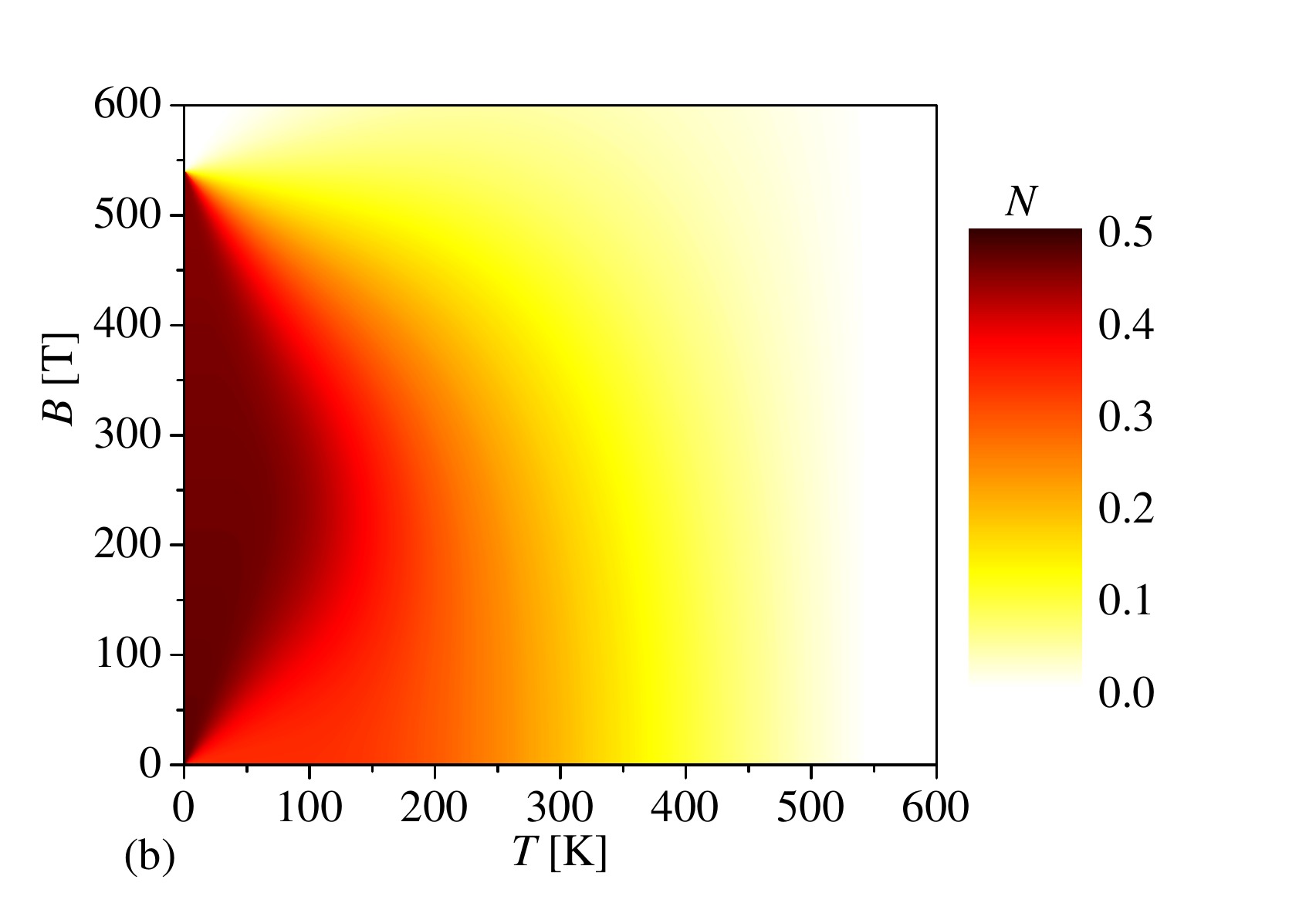}
\end{center}
\vspace*{-0.8cm}
\caption{Density plots of the negativity in the temperature versus magnetic field plane for the mixed spin-($\frac{1}{2}$,1) Heisenberg dimer given by the Hamiltonian (\ref{eq1}) with the parameter set $J/k_B = 505$~K, $g_{Rad} = 2.005$, and $g_{Ni} = 2.275$ adjusted to a theoretical modeling of the molecular complex (Et$_3$NH)[Ni(hfac)$_2$L]. Fig. \ref{fig4}(a) shows a detail of the density plot for the experimentally accessible magnetic-field range 0--10~T (note that there is an axis break between a low- and high-temperature region), while Fig. \ref{fig4}(b) shows a complete density plot up to magnetic fields above which the negativity vanishes.}
\label{fig4}
\end{figure}

Finally, let us conclude our discussion with a detailed analysis of a density plot of the negativity in the temperature versus magnetic field plane, which was obtained for the mixed spin-($\frac{1}{2}$,1) Heisenberg dimer with the parameter set $J/k_B = 505$~K, $g_{Rad} = 2.005$, $g_{Ni} = 2.275$ adjusted for a theoretical modeling of the molecular complex (Et$_{3}$NH)[Ni(hfac)$_{2}$L].  Fig. \ref{fig4}(a) shows a detail of the relevant density plot up to the magnetic field $B = 10$~T, which represents a typical upper limit of the magnetic-field strength for commercially accessible devices generating static magnetic fields. It is obvious from a low-temperature part of the density plot that this magnetic-field range is capable of stabilizing almost perfect bipartite entanglement ${\cal N} \approx 0.47$, which nearly approaches fully entangled Bell states, only up to temperature $T \approx 5$~K. In fact, the rise in temperature necessitates a stronger magnetic field to restore almost perfect bipartite  entanglement as reflected by a relatively steep gradient of the dark region in the given density plot. A color shading with character of vertical stripes observed in a high-temperature part of the density plot ($T > 50$~K) contrarily verifies the independence of the strength of  bipartite thermal entanglement on the magnetic field in the range of sufficiently small magnetic fields $B < 10$~T. Although the strengthening of bipartite thermal entanglement due to the Zeeman effect cannot be utilized in the high-temperature region, the increasing magnetic field does not diminish the strength of bipartite thermal entanglement either. For completeness, the density plot of the negativity in a full range of temperatures and magnetic fields is plotted in Fig. \ref{fig4}(b) in order to provide the phase diagram, which delimits a parameter region with nonzero thermal entanglement. The displayed phase diagram convincingly evidences that the molecular compound (Et$_{3}$NH)[Ni(hfac)$_{2}$L] exhibits the thermal entanglement, which is confined to temperatures below $540$~K and magnetic fields below $540$~T. The respective limit set for temperature evidently implies a possible implementation of the nickel-radical molecular complex (Et$_{3}$NH)[Ni(hfac)$_{2}$L] for room-temperature technological applications exploiting the intense thermal entanglement for quantum computation and quantum information processing. 

\section{Conclusion}
\label{sec:conc}

In this paper, we have adapted exact results derived for the negativity of the mixed spin-($\frac{1}{2}$,1) Heisenberg dimer to gain insight into the strength of bipartite quantum and thermal entanglement of the mononuclear molecular complex (Et$_{3}$NH)[Ni(hfac)$_{2}$L], which is characterized by a sizable exchange coupling constant between the spin-$\frac{1}{2}$ nitronyl-nitroxide radical substituted nitrophenol and the spin-1 Ni$^{2+}$ magnetic ion \cite{spin21}. The significant energy gap $\Delta E/k_B = 758$~K between the quantum ferrimagnetic ground-state doublet and a quartet excited state suggests that the molecular complex (Et$_{3}$NH)[Ni(hfac)$_{2}$L] provides a suitable platform of a molecular qubit in a wide range of temperature and magnetic fields. It has been demonstrated that the Zeeman splitting of the two-fold degenerate quantum ferrimagnetic ground state significantly reinforces the strength of  bipartite entanglement at small magnitudes of the magnetic field, whereas a further enhancement of the magnetic field gradually suppresses the strength of bipartite entanglement. The latter reduction of the bipartite entanglement is however quite weak effect for achievable magnetic-field strengths, because it is inhibited by a competition of the extremely strong exchange coupling constant $J/k_B = 505$~K relative to the magnetic-field-dependent term $(g_{Ni}-g_{Rad}) \mu_B B$ involving a rather small difference between g-factors g$_{Rad}$=2.005 and g$_{Ni}$=2.275 of the spin-$\frac{1}{2}$ nitronyl-nitroxide radical and the spin-1 Ni$^{2+}$ magnetic ion, respectively. It could be thus concluded that the nickel-radical molecular complex (Et$_{3}$NH)[Ni(hfac)$_{2}$L] affords a prominent example of the molecular qubit whose entanglement strength can be substantially enhanced (cca. 41\%) by a small driving magnetic field. This significant strengthening of bipartite entanglement in response to the application of small magnetic fields $B < 10$~T indeed persists in the molecular complex (Et$_{3}$NH)[Ni(hfac)$_{2}$L] up to temperatures $T \lesssim 5$~K highlighting the potential significance of this magnetic compound for experimental verification of this remarkable phenomenon predicted for experimental realizations of the mixed spin-$(\frac{1}{2}, S)$ Heisenberg dimers \cite{cenc20,varg21}. 

Finally, it is worthwhile to remark that the molecular compound (Et$_{3}$NH)[Ni(hfac)$_{2}$L] maintains even at room temperature a sufficiently strong bipartite entanglement retaining approximately 40\% of the maximum value corresponding to perfectly entangled Bell states. This intriguing feature persists even at extremely high magnetic fields considerably exceeding commercially available static magnetic fields. From this perspective, the molecular complex (Et$_{3}$NH)[Ni(hfac)$_{2}$L] can be regarded as a compelling example of a molecular qubit, which may be implemented for prospective technological applications exploiting quantum computing and quantum information processing at room temperature including data storage devices and quantum gates. Our further goal is therefore to examine the relaxation times and quantum spin dynamics of this molecular complex.

	\vspace{6pt} 
	
	\authorcontributions{Conceptualization, J.S.; methodology, J.S.; software, J.S. and E.S.S.; validation, J.S., and E.S.S.; formal analysis, J.S., and E.S.S.; investigation, J.S., and E.S.S.; resources, J.S.; data curation, J.S. and E.S.S.; writing---original draft preparation, J.S. and E.S.S.; writing---review and editing, J.S. and E.S.S.; visualization, J.S. and E.S.S.; supervision, J.S.; project administration, J.S.; funding acquisition, J.S. All authors have read and agreed to the published version of the manuscript.}
	
	\funding{This research was funded by Ministry of Education, Science, Research and Sport of the Slovak Republic under the grant number 
		VEGA 1/0695/23 and by Slovak Research and Development Agency under the grant number APVV-20-0150.}
	
	\institutionalreview{Not applicable.}
	
	\informedconsent{Not applicable.}
	
	\dataavailability{Not applicable.} 
	
	\conflictsofinterest{The authors declare no conflict of interest.} 
	
	\abbreviations{Abbreviations}{The following abbreviations are used in this manuscript:\\
		HL = 2-(2-hydroxy-3-methoxy-5-nitrophenyl)-4,4,5,5-tetramethyl-4,5-dihydro-1H-imidazol-3-oxide-1-oxyl\\ 
		hfacH = hexafluoroacetylacetone\\
		Et = ethyl}
	
	\appendixtitles{no} 
	\appendixstart
	\appendix
	\section{}
	\label{appendixa}
	The explicit form of non-zero elements $\rho_{ij}$ of the density matrix given by equation~(\ref{eq12}). 
	\allowdisplaybreaks

	\begin{adjustwidth}{-\extralength}{0cm}
		\begin{align}
		\rho_{11}&=\frac{1}{Z} \exp \left\{-\frac{\beta}{2} \left[J-\left(h_{1}+2h_2\right)\right] \right\};
		\label{eqA1}\\
		\rho_{22}&=\frac{1}{Z} \exp\left[\frac{\beta}{4} \left(J+2h_{2}\right) \right] \left[ \cosh\left(\frac{\beta}{4}\sqrt{\left[J\!-\!2(h_1\!-\!h_2)\right]^2\!+\!8J^2}\right) \right.
		\nonumber\\ &- \left. \frac{J\!-\!2(h_1\!-\!h_2)}{\sqrt{\left[J\!-\!2(h_1\!-\!h_2)\right]^2\!+\!8J^2}} \right.
		\left. \sinh\left(\frac{\beta}{4}\sqrt{\left[J\!-\!2(h_1\!-\!h_2)\right]^2\!+\!8J^2}\right) \right];
		\label{eqA2}\\
		\rho_{33}&=\frac{1}{Z} \exp\left[\frac{\beta}{4} \left(J-2h_{2}\right) \right] \left[ \cosh\left(\frac{\beta}{4}\sqrt{\left[J\!+\!2(h_1\!-\!h_2)\right]^2\!+\!8J^2}\right) \right.
		\nonumber\\ &+ \left.
		\frac{J\!+\!2(h_1\!-\!h_2)}{\sqrt{\left[J\!+\!2(h_1\!-\!h_2)\right]^2\!+\!8J^2}}\right.
		\left.\sinh\left(\frac{\beta}{4}\sqrt{\left[J\!+\!2(h_1\!-\!h_2)\right]^2\!+\!8J^2}\right) \right];
		\label{eqA4}\\
		\rho_{44}&=\frac{1}{Z} \exp\left[\frac{\beta}{4}\left(J+2h_{2}\right) \right] \left[ \cosh\left(\frac{\beta}{4}\sqrt{\left[J\!-\!2(h_1\!-\!h_2)\right]^2\!+\!8J^2}\right) \right.
		\nonumber\\ &+ \left.
		\frac{J\!-\!2(h_1\!-\!h_2)}{\sqrt{\left[J\!-\!2(h_1\!-\!h_2)\right]^2\!+\!8J^2}}\right.
		\left.\sinh\left(\frac{\beta}{4}\sqrt{\left[J\!-\!2(h_1\!-\!h_2)\right]^2\!+\!8J^2}\right) \right];
		\label{eqA6}\\
		\rho_{55}&=\frac{1}{Z} \exp\left[\frac{\beta}{4} \left(J-2h_{2}\right) \right] \left[ \cosh\left(\frac{\beta}{4}\sqrt{\left[J\!+\!2(h_1\!-\!h_2)\right]^2\!+\!8J^2}\right) \right.
		\nonumber\\ &+ \left.
		-\frac{J\!+\!2(h_1\!-\!h_2)}{\sqrt{\left[J\!+\!2(h_1\!-\!h_2)\right]^2\!+\!8J^2}}\right.
		\left.\sinh\left(\frac{\beta}{4}\sqrt{\left[J\!+\!2(h_1\!-\!h_2)\right]^2\!+\!8J^2}\right) \right];
		\label{eqA7}\\
		\rho_{66}&=\frac{1}{Z} \exp\left\{-\frac{\beta}{2} \left[J+\left(h_{1}+2h_2\right)\right] \right\};
		\label{eqA8}\\
		\rho_{24}&=\rho_{42}\!=\!-\frac{\sqrt{8}J \exp\left[\frac{\beta}{4} \left(J+2h_{2}\right) \right]}{Z\sqrt{\left[J\!-\!2(h_1\!-\!h_2)\right]^2\!+\!8J^2}}\sinh\left(\frac{\beta}{4}\sqrt{\left[J\!-\!2(h_1\!-\!h_2)\right]^2\!+\!8J^2}\right);
		\label{eqA3}\\
		\rho_{35}&=\rho_{53}\!=\!-\frac{\sqrt{8}J \exp\left[\frac{\beta}{4} \left(J-2h_{2}\right)\right]}{Z\sqrt{\left[J\!+\!2(h_1\!-\!h_2)\right]^2\!+\!8J^2}}\sinh\left(\frac{\beta}{4}\sqrt{\left[J\!+\!2(h_1\!-\!h_2)\right]^2\!+\!8J^2}\right).
		\label{eqA5}
		\end{align}
	\end{adjustwidth}

	\begin{adjustwidth}{-\extralength}{0cm}
				
		\reftitle{References}

	\end{adjustwidth}
	
\end{document}